\documentstyle[twoside,fleqn,espcrc2,epsfig]{article}

\newcommand{\beq}{\begin{equation}}
\newcommand{\eeq}{\end{equation}}
\newcommand{\bea}{\begin{eqnarray}}
\newcommand{\eea}{\end{eqnarray}}
\newcommand{\gsim}{\raisebox{-0.05cm}{$\:\stackrel{>}{{\scriptstyle
 \sim}}\: $} }
\newcommand{\lsim}{\raisebox{-0.05cm}{$\:\stackrel{<}{{\scriptstyle
 \sim}}\: $} }

\title{Parton densities and structure functions beyond the 
       next-to-leading order}

\author{W.L. van Neerven and A. Vogt
        \address{Instituut-Lorentz, University of Leiden,
        P.O. Box 9506, 2300 RA Leiden, The Netherlands}%
        \thanks{Work supported by the EC TMR program under contract
        No.\ FMRX--CT98--0194. Presented by A. Vogt at `Loops and 
        Legs in QFT', Bastei (Germany), April 2000.} }

\begin{document}

\begin{abstract}
We discuss recent results on the evolution of unpolarized parton 
densities and structure functions in massless perturbative QCD.  
Present partial results on the next-to-next-to-leading order (NNLO)
evolution kernels prove sufficient for reliable calculations at not too 
small values  of the Bjorken variable, $x\! >\! 10^{-3}$.  One order 
more can be taken into account at $x\!\geq\! 0.2$.  Inclusion of these 
terms considerably reduces the main theoretical uncertainties of 
determinations of $\alpha_s$ (to about 1\% at the $Z$-mass) and the 
parton densities from structure functions.
\end{abstract}

\maketitle

\section{Introduction}

Structure functions in deep-inelastic scattering (DIS) are among the 
quantities best suited for measuring the strong coupling constant 
$\alpha_s$.  They also form the backbone of our knowledge of the 
proton's parton densities, indispensable for analyses of hard 
scattering processes at proton--(anti-)proton colliders like 
{\sc Tevatron} and the LHC.  During the past two decades DIS 
experiments have proceeded towards high accuracy and a greatly extended 
kinematic coverage~\cite{exp}.

\vspace{1mm}
To make full use of these results requires transcending the standard 
next-to-leading order (NLO) formalism \cite{FP82}. Indeed besides the 
QCD $\beta$-functions to even NNNLO \cite{beta2,beta3}, the NNLO 
\mbox{(2-loop)} coefficient functions for DIS have been calculated some 
time ago \cite{ZvN1,ZvN2}. However, only partial results have been 
obtained so far for the corresponding 3-loop splitting functions 
\cite{spfm1,spfm2,Gra1,BVns,Gra2,CH94,FL98}.  The derivation of the 
full results is under way~\cite{MV2}.

\vspace{1mm}
In \cite{NV1,NV2} we have derived approximate expressions for the 
3-loop $\overline{\mbox{MS}}$ splitting functions which are sufficient 
for reliable NNLO analyses down to  $x \!\simeq\! 10^{-3}$.  These 
\mbox{functions} turn out to be much less important than the 2-loop 
coefficient functions at $x\!\geq\! 10^{-2}$. Thus it is possible, 
based on partial results \cite{spfm1,spfm2,sglue,av99} on the 3-loop 
coefficient functions, to proceed to NNNLO at large $x$ \cite{NV3}, 
especially for the non-singlet case most important for extractions of 
$\alpha_s$ from DIS. 

\section{Parton densities: formalism}

It is convenient to work with the flavour non-singlet (NS) and singlet 
(S) combinations of the (anti-)quark and gluon densities, $q_i$, 
$\bar{q}_i$ and $g\,$:
\bea
\label{ave1}
  q_{\rm NS}^{\,\pm} &\!\! =\!\! & q_i\pm\bar{q}_i - (q_k\pm \bar{q}_k)
  % \:\: , \:\: i \neq k = 1, \ldots, N_f 
  \nonumber \\
  q_{\rm NS}^V     &\!\! =\!\! & \sum_{r=1}^{N_f} (q_r - \bar{q}_r) 
  \\[-1mm]
  q_{S}^{\,}\:     &\!\! =\!\! & \left( \begin{array}{c} \!\Sigma\! \\
  \! g\! \end{array} \right) \:\: , \:\:
  \Sigma \: = \:\sum_{r=1}^{N_f} (q_r + \bar{q}_r) \:\: .
  \nonumber
\eea
Here $N_f$ is the number of effectively massless flavours. As in 
(\ref{ave1}) we often suppress the dependence on the momentum fraction 
$x$ and the renormalization and factorization scales, $\mu_r$ 
and~$\mu_f$.

\vspace{1.5mm}
Using (\ref{ave1}) the evolution equations are decomposed into
$2N_f\! -\! 1$ scalar (NS) equations and the $2\times 2$ singlet 
system, all schematically written as
\beq
\label{ave2}
  \frac{d}{d \ln \mu_f^2} \: q = \: {\cal P} \otimes q \:\equiv\: 
  \int_x^1 \! \frac{dy}{y}\, {\cal P}(y)\, q\bigg(\frac{x}{y}\bigg)
  \:\: . 
\eeq
At $\mu_r\! =\! \mu_f$ the expansion of the splitting functions 
${\cal P}$ up to NNLO is given by
\beq
\label{ave3}
   {\cal P} \:\simeq\: a_s \, P^{(0)}(x) + a_s^2 \, P^{(1)}(x)
   + a_s^3 \, P^{(2)}(x) \:\: . 
\eeq
Our choice of the expansion parameter reads
\beq
\label{ave4}
  a_s \:\equiv\: \alpha_s/4\pi \:\: .
\eeq
The expression for $\mu_r\! \neq\! \mu_f$ is obtained from (\ref{ave3}) 
by inserting the expansion of $a_s(\mu_f^2)$ in terms of $a_s(\mu_r^2)$
and $L_R = \ln (\mu_f^2/\mu_r^2)$. Large logarithms in ${\cal P}$ are 
avoided by choosing $\mu_r = O(\mu_f)$. 
 
\section{Splitting functions}
 
$\!\!$The functions $P^{(0)}\,$and $P^{(1)}$ in (\ref{ave3}) are 
known~\cite{FP82}. The current information on $P_{\rm NS}^{(2)\pm}(x)$ 
comprises
\begin{itemize}
\item the first five even-integer moments of $P^{(2)+}_{\rm NS}$ 
      given by $P^{(2)+}_{\rm NS}(N) = \int_0^1 \! dx\: x^{N-1} 
      P^{(2)+}_{\rm NS}(x)$ \cite{spfm1,spfm2}, and the first moment 
      ($N$=1) of $P^{(2)-}_{\rm NS}$,
\item the complete $N_f^2$ contribution \cite{Gra1},
\item the leading small-$x$ terms $\propto \ln ^4 x$ \cite{BVns}.
\end{itemize}
The difference of $P^{(2)-}_{\rm NS}$ and $P^{(2)+}_{\rm NS}$
is expected to be negligible at large $x$. It has been conjectured
that the leading large-$x$ terms are $\propto\! 1/[1-x]_+$\cite{GLY}.   

\vspace{1.5mm}
The following partial results have been derived so far for the 
singlet splitting functions $P^{(2)}_{ij}(x)$:
\begin{itemize}
\item $P^{(2)}_{ij}(N)$ for $N = 2,\, 4,\, 6$ and 8 \cite{spfm2},
\item the $C_A N_f^2$ contribution to $P^{(2)}_{gg}(x)$ \cite{Gra2},
\item the leading small-$x$ terms $\propto (1/x) \,\ln x$ of
      $P^{(2)}_{qq}\!$, $P^{(2)}_{qg}$ \cite{CH94} and $P^{(2)}_{gg}$
      \cite{FL98}, see also \cite{BNRV}. 
\end{itemize}
The $1/[1-x]_+$ terms of $P^{(2)}_{gg}$ and $P^{(2)}_{qq}$ are expected 
to be related by a factor $C_A/C_F = 9/4$.

\vspace{1.5mm}
We have derived approximate expressions for $P_{\rm NS}^{(2)\pm}(x)$ 
and $P^{(2)}_{ij}(x)$ from these constraints. 
After decomposing the functions into
\beq
\label{ave5}
  P^{(2)} \: =\:  P_{0}^{(2)} + N_f P_{1}^{(2)}
  + N_f^2 P_{2}^{(2)} \:\: ,
\eeq
we employ the ansatz (cf.~\cite{spfm2})
\beq
\label{ave6}
   P_{m}^{(2)}(x) \: =\: \sum_{n=1}^{n_m} A_n f_n(x) + f_e(x) \:\: .
\eeq
The basis functions $f_n$ are build up of $1/[1-x]_+$, $\delta (1-x)$ 
and of powers of $\ln (1-x)$, $x$, and $\ln\, x$. The coefficients 
$A_n$ are determined from the $n_m\! =\! 5$ ($n_m\! =\! 4$) linear 
equations provided by the non-singlet (singlet) moments of 
\cite{spfm1,spfm2} after taking into account the other constraints 
collected in $f_e$ in~(\ref{ave6}).  The remaining uncertainties are 
estimated by `reasonably' varying the choice of the basis functions 
$f_n$, typically considering some 20 to 40 trial functions.  Finally 
two approximations spanning the error band are selected, except for the 
highest unknown $N_f$-contributions in (\ref{ave5}) for which one 
central representative is sufficient.

\begin{figure}[htb]
\vspace{-6mm}
\centerline{\epsfig{file=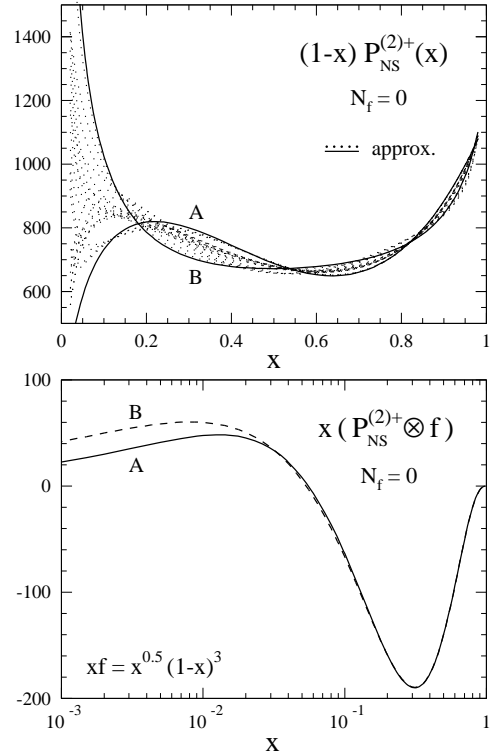,width=6.6cm,angle=0}}
\vspace{-1cm}
\caption{Top: $P_{\rm NS}^{(2)+}$ for $N_f\! =\! 0$, as obtained from 
 the results of \cite{spfm1,spfm2,BVns} by means of (\ref{ave6}). 
 Bottom: Convolutions of the selected approximations A and B with a 
 typical non-singlet input shape.}
\vspace*{-0.5mm}
\label{avf1}
\end{figure}

\begin{figure}[htb]
\centerline{\epsfig{file=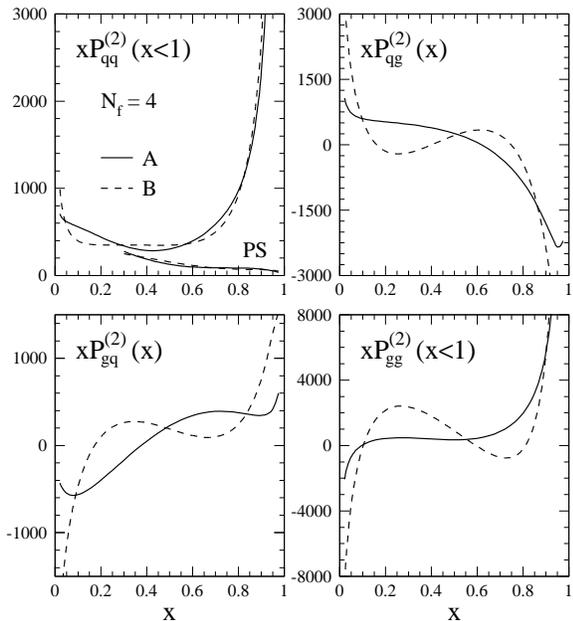,width=7.7cm,angle=0}}
\vspace{-8mm}
\caption{The approximations selected for $P_{ij}^{(2)}$ for $N_f\! =\! 
 4$.  The pure singlet (PS) contribution $P_{qq}^{(2)} - 
 P_{\rm NS}^{(2)+}$ is also shown at $x \geq 0.3$.}
\vspace{-4mm}
\label{avf2}
\end{figure}

\vspace{1mm}
This procedure is briefly illustrated in Fig.~\ref{avf1} for the
$N_f\, =0\, $ part of $P_{\rm NS}^{(2)+}(x)$.  The upper plot shows 24 
trial functions.  The approximations A and B emphasized in the plot
have been selected, after considering also the convolution with a 
typical input shape shown for these two functions in the lower plot.
As can be inferred from Fig.~\ref{avf1}, the presence of the 
convolution in (\ref{ave2}) considerably increases the effective 
accuracy of our approximations illustrated in Figs.~\ref{avf1} and 
\ref{avf2}: The convolutions smoothen out the oscillating 
\mbox{large-$x$} differences between different approximations to a 
large extent.  They also partly compensate the large small-$x$ 
uncertainties of $P^{(2)}$ present despite the $x\!\rightarrow\! 0$
 constraints of \cite{BVns,CH94,FL98}.

\section{Parton densities: results}

We illustrate the impact of the NNLO terms on the parton evolution by 
the derivatives $\dot{q} \equiv d\ln q / d\ln \mu_f^2$,  $q = 
q_{\rm NS}^{\, +},\, \Sigma,\, g$, at a reference scale
\beq
\label{ave7}
  \mu_f^2 \: =\: \mu_{f,0}^2 \:\approx\: 30 \mbox{ GeV}^2
\eeq
corresponding to $\alpha_s (\mu_r^2\! =\! \mu_{f,0}^2) = 0.2$.
The input densities adopted for the non-singlet case read
\beq
\label{ave8}
  xq_{\rm NS}^{\, +} \: = \: x^{0.5} (1-x)^3 \:\: .
\eeq
For the singlet distributions we employ
\bea
\label{ave9}
  x\Sigma (x,\mu_{f,0}^2) &\!\! =\!\! & 0.6\, x^{-0.3}\, (1-x)^{3.5}
                                        (1 + 5\, x^{0.8}) 
  \nonumber \\
  xg (x,\mu_{f,0}^2) &\!\! =\!\! & 1.0\, x^{-0.37} (1-x)^{5} \:\: .
\eea
The dependence of the results on the renormalization scale is presented 
via
\beq
\label{ave10}
  \Delta \dot{q} \:\equiv\: \frac{\max\, \dot{q} - \min\, \dot{q} }
  { 2\, |\, {\rm average}\, \dot{q}\, | } \:\: , 
\eeq
where $\mu_r$ is varied over the conventional interval
\beq
\label{ave11}
  1/4\: \mu_{f}^2 \:\leq\: \mu_r^2 \:\leq\: 4\, \mu_{f}^2 \:\: .
\eeq
 
\begin{figure}[htb]
\vspace{-7mm}
\centerline{\epsfig{file=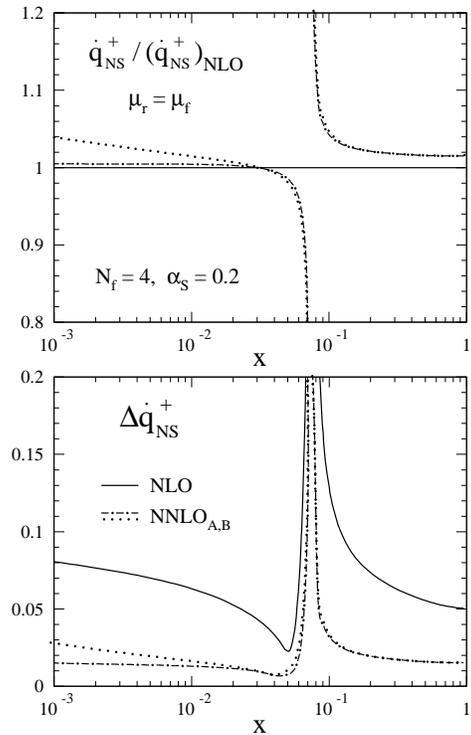,width=6.4cm,angle=0}}
\vspace{-1cm}
\caption{Top: The NNLO corrections for $\dot{q}_{\rm NS}^{\, +} \equiv 
 d\ln q_{\rm NS}^{\, +}/ d\ln \mu_f^2$ at $\mu_f\! = \mu_{f,0}$ for the
 input (\ref{ave8}).
 Bottom: The relative $\mu_r$-uncertainty of the NLO and NNLO results
 for $\dot{q}_{\rm NS}^{\, +}$ using (\ref{ave10}) and (\ref{ave11}).}
\vspace*{-1mm}
\label{avf3}
\end{figure}

\begin{figure}[htb]
\centerline{\epsfig{file=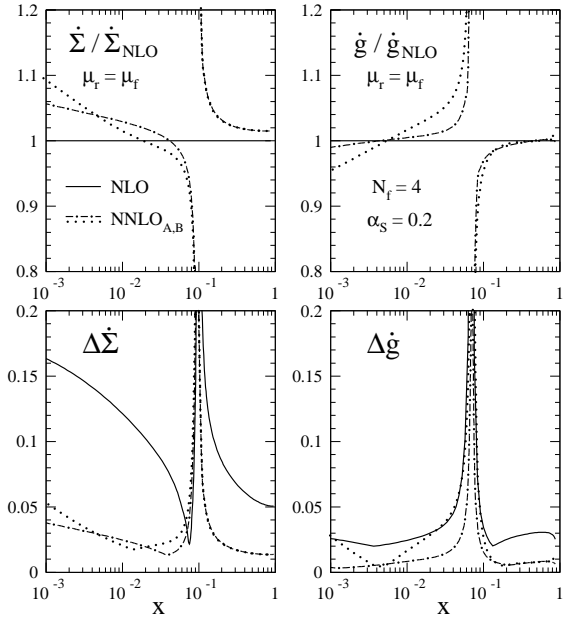,width=7.7cm,angle=0}}
\vspace{-8mm}
\caption{As Figure~\ref{avf3}, but for the singlet quark and gluon 
 densities $\Sigma$ and $g$ given in (\ref{ave8}) and (\ref{ave9}).
 The spikes close to $x\! =\! 0.1$ in both figures are due to zeros 
 of the respective denominators.}
\vspace{-4mm}
\label{avf4}
\end{figure}

The NNLO effects on the derivatives $\dot{q}$ and the NLO and NNLO 
scale uncertainties $\Delta \dot{q}$ are shown in Figs.~\ref{avf3} and 
\ref{avf4}.  The present inaccuracies of the NNLO results caused by 
the uncertainties remaining for the functions $P^{(2)}$ are represented 
by the bands spanned by the NNLO$_A$ and NNLO$_B$ curves.  The central 
results $\frac{1}{2}(\mbox{NNLO}_A + \mbox{NNLO}_B$) are not shown 
separately.

\vspace{1.5mm}
The uncertainties of the NNLO derivatives $\dot{q}$ due to the 
approximations for $P^{(2)}$ are entirely negligible for $x \gsim 0.1$. 
They increase towards very small values of $x$, but do not exceed 
$\pm 2\%$ above $x \simeq 10^{-3}$ (or a few times this number for 
scales $\mu_f$ much smaller than (\ref{ave7})$\,\!$).  Given the small 
size of the NNLO corrections and the weak $\mu_r$-dependence remaining
at NNLO, one can safely estimate that contributions beyond NNLO affect 
the parton evolution, for $\alpha_s \lsim 0.2$, by less than 1\% at 
large $x$ and 2\% down to $x\simeq 10^{-3}$.

\section{Structure functions: formalism}

The unpolarized non-singlet ($a=1,\, 2,\, 3$) and singlet ($a=1,\, 2$) 
structure functions $F_a$ are obtained by convoluting the solutions 
$q(\mu_f^2, \mu_r^2)$ of (\ref{ave2}) with the corresponding 
coefficient functions, 
\bea
\label{ave12}
 \eta_a F_a(x,Q^2) \: = \: \big[ \,{\cal C}_a(a_s, L_M, L_R) 
 \otimes q\, \big] (x) \:\: .
\eea
For $L_M \! = \ln\, (Q^2/\mu_f^2)\!\neq\! 0$, but $\mu_r\! 
= \mu_f$ ($L_R\! = 0)\,$:
\bea
\label{ave13}
  & & \nonumber \\[-5mm]
  {\cal C}_a \!\!\!\! & = & \!\!\! c_a^{(0)}(x) + \sum_{l=1} a_s^l 
  \Big\{ c_a^{(l)}(x) + \!\sum_{m=1}^{l}\! c_a^{(l,m)}(x)\, L_M^m 
  \!\Big\} \nonumber \\[-2mm] & & 
\eea
with ${\cal C}_{a}\! = (\, {\cal C}_{a,q}\, , \, {\cal C}_{a,g}\, )$ 
in the singlet case. The coefficients $\eta_a$ in (\ref{ave13}) include 
the charge factors so that $c_{a,\rm NS}^{(0)}\! =\!  c_{a,q}^{(0)}\! 
=\! \delta (1\! -\! x)$, whereas, of~course,\\[-0.5mm]
$c_{a,g}^{(0)}\! =\! 0$. 
The contributions $c_a^{(l,m)}$ fixed by renormalization-group 
constraints are build up of the $c_a^{(k)}$ and the splitting functions 
$P^{(k)}$ up to $k = l\! -\! 1$.  The generalization of (\ref{ave13}) 
to $\mu_r\neq \mu_f$ proceeds as indicated below (\ref{ave4}).

\vspace{1.5mm}
The scaling violations of the non-singlet structure functions can be 
conveniently expressed in terms of these structure functions themselves 
(thereby removing any dependence on $\mu_f$), viz 
\beq
\label{ave14}
  \frac{d}{d \ln Q^2} \, F_{a,\rm NS} \: = \:
  {\cal K}_{a,\rm NS} \otimes F_{a,\rm NS} \:\: .
\eeq
The kernels ${\cal K}_{a,\rm NS}$ are derived by differentiating
(\ref{ave12}) with respect to $\ln Q^2$ and then eliminating the 
quark densities using the same equation.

\vspace{2mm}
\section{Coefficient functions}

Besides the functions $c_{a}^{(1)}$ in (\ref{ave13}), see \cite{FP82},
also the NNLO contribution $c_{a}^{(2)}$ are known \cite{ZvN1,ZvN2}.
Those expressions are rather lengthy and involve higher 
transcendental functions. We have thus provided compact 
approximations which are sufficiently accurate for any foreseeable
application.

\vspace{1.5mm}
As illustrated below (Fig.~6), the impact of the functions 
$c_{a}^{(2)}$ (especially of the quark coefficient functions which 
contain large soft-gluon emission terms) is much larger than that of 
the splitting functions $P^{(2)}$ at $x>10^{-2}$. The same situation is 
expected for the NNNLO quantities $c_{a}^{(3)}$ and $P^{(3)}$. Hence a 
good approximation to the NNNLO at large $x$ can be obtained by just 
retaining the $c_{a}^{(3)}$.

\vspace{1.5mm}
The current information on the $c_{a}^{(3)}$ comprises
\begin{itemize}
\item the first five even-integer moments of $c_{2,\rm NS}^{(3)+}$,
      and the first four of $c_{2,q}^{(3)}$ and $c_{2,g}^{(3)}$
      \cite{spfm1,spfm2},
\item the four leading large-$x$ terms $\propto \ln^k (1\! -\! x)/$
      $[1-x]_+$, $k = 2,\,\ldots ,\, 5$ of $c_{a,\rm NS}^{(3)}$ and 
      $c_{a,q}^{(3)}$ \cite{spfm2,av99}, fixed by the results of 
      \cite{sglue}, together with those of \cite{ZvN1} for $k = 2$.
\end{itemize}
For $c_{2,\rm NS}^{(3)-}$ only the first moment (= 0) is known from the 
Adler sum rule. However, the results on $c_{2,\rm NS}^{(2)\pm}$ 
indicate that the difference $c_{2,\rm NS}^{(3)-}-c_{2,\rm NS}^{(3)+}$ 
has a negligibly small effect. Results for the lowest moments of 
$c_{\, 3}^{(3)}$ will become available soon \cite{Jos}.

\vspace{1.5mm}
Focusing on the non-singlet case most relevant for $\alpha_s$-% 
determinations from structure functions, we have employed the above 
information to derive approximations of $c_{2,\rm NS}^{(3)}(x)$ 
\cite{NV3}.  The impact of their residual uncertainties is small for 
$x \geq 0.2$. 

\section{Structure functions: results}

We illustrate the effect of the NNLO (and NNNLO) terms on the structure 
functions at
\beq
\label{ave15}
  Q^2 \:\approx\: 30 \mbox{ GeV}^2 \:\: .
\eeq
In (\ref{ave14}) we employ the non-singlet input shape
\beq
\label{ave16}
  F_{2,\rm NS}^{\, +}(x,Q^2) \: = \: x^{0.5} (1-x)^3 \:\: .
\eeq
For the singlet case we fix, besides $\alpha_s (Q^2) = 0.2$, the 
input parton densities (\ref{ave9}), hence not $F_{2,S}$.  The 
$\mu_r$-dependence of the results for $f= F_{2,S}$, $\dot{F}_{2,\rm NS} 
\equiv d\ln F_{2,\rm NS}/ d\ln Q^2$, $\dot{F}_{2, S}$ and 
$F_{2,S}^{\,\prime} \equiv dF_{2,S}/ d\ln Q^2$ is presented via
\beq
\label{ave17}
  \Delta \dot{f} \equiv \frac{\max\, \dot{f} - \min\, \dot{f} }
  { 2\, |\, {\rm average}\, \dot{f}\, | } \:\: , \:\:\:
  \frac{1}{4}\, Q^2 \leq \mu_r^2 \leq 4\, Q^2 \:\: . \quad
\eeq
The singlet results are shown for $\mu_f = \mu_r \equiv \mu$.
 
\begin{figure}[htb]
\vspace{-6mm}
\centerline{\epsfig{file=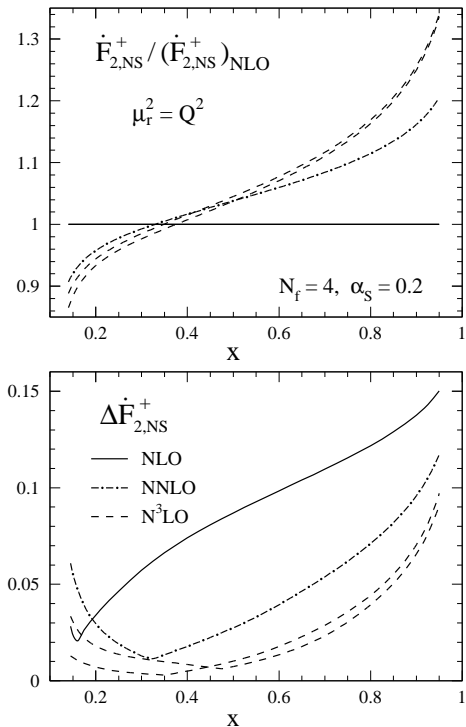,width=6.4cm,angle=0}}
\vspace{-1cm}
\caption{Top: The NNLO and NNNLO corrections for $\dot{F}_{2,\rm NS}
 ^{\, +}$ for the input (\ref{ave15}) and (\ref{ave16}). 
 Bottom: The relative $\mu_r$-uncertainties (\ref{ave17}) compared to 
 the corresponding NLO result.}
\vspace*{-1mm}
\label{avf5}
\end{figure}

The results for $\dot{F}_{2,\rm NS}$ are shown in Fig.~\ref{avf5}. The 
uncertainty bands for the NNNLO predictions take into account both the
remaining inaccuracies of the coefficient functions 
$c_{2,\rm NS}^{(3)}(x)$ and the possible effects of the splitting 
functions $P_{\rm NS}^{(3)}$.  At $0.25 \lsim x \lsim 0.7$ the 
$\mu_r$-uncertainties $\Delta \dot{F}_{2,\rm NS}$ are reduced by a 
factor of two (four) or more at NNLO (NNNLO). These uncertainties lead 
to the following estimates for the errors of $\alpha_s(M_Z^2)$ due to 
the truncation of the perturbation series:
\bea
  \Delta \alpha_s (M_Z^2)_{\rm NLO}\:\:\:\:\,  & \simeq & \! 0.005 
  \nonumber \\
  \Delta \alpha_s (M_Z^2)_{\rm NNLO}\:\: & \simeq & \! 0.002 
  \\
  \Delta \alpha_s (M_Z^2)_{\rm NNNLO}\!  & \simeq & \! 0.001 \:\: .
  \nonumber 
\eea
A 1\% accuracy is achieved at the NNNLO level.
As the scaling violations for the same $\alpha_s(Q^2)$ are stronger
at NNLO and NNNLO than at NLO, higher-order fits of data on 
$F_{2,\rm NS}$ will yield somewhat lower central values of 
$\alpha_s(M_Z^2)$,
\beq
 \alpha_s(M_Z^2)_{\rm NNNLO} - \alpha_s(M_Z^2)_{\rm NLO} \:\approx\:
 -\, 0.002 \:\: .
\eeq

\begin{figure}[htb]
\centerline{\epsfig{file=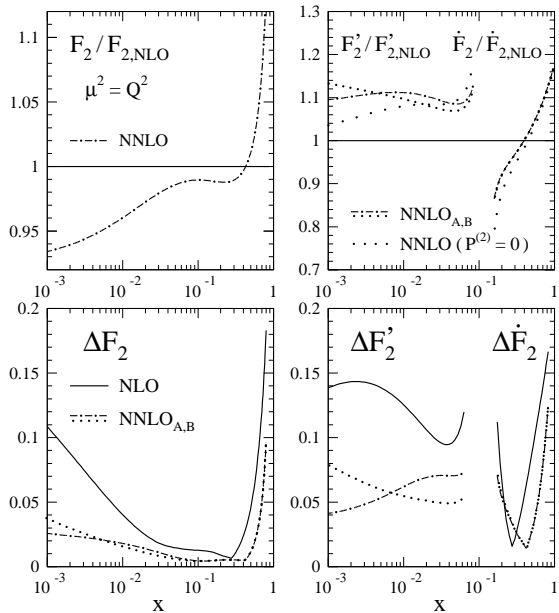,width=7.6cm,angle=0}}
\vspace{-9mm}
\caption{Top: The NNLO corrections for $F_2\equiv F_{2.S}$ and its 
 $Q^2$ derivatives for the input (\ref{ave9}) at $\mu_{f,0}^2\! =\! 
 Q^2$.
 Bottom: The corresponding $\mu_r\! =\! \mu_f$ uncertainties 
 (\ref{ave17}) compared to the NLO results.}
\vspace{-6mm}
\label{avf6}
\end{figure}

\vspace{1mm}
The results for the singlet case are presented in Fig.~\ref{avf6}.
$F_{2,S}$ receives large positive corrections at large $x$, caused by
the soft-gluon parts of the quark coefficient functions. The sizeable 
negative NNLO corrections at small $x$ are dominated by the gluon 
contribution.  It is worth noting that the positive $1/x$ term of 
$c_{2,g}^{(2)}$ does not dominate this correction even at $x< 10^{-3}$.

\vspace{1mm}
The $Q^2$-derivative of $F_{2,S}$ is dominated by the quark 
contribution at $x\! >\! 0.3$, and by the gluon contribution at 
$x\! <\! 0.03$.  Thus we present the logarithmic derivative 
$\dot{F}_{2,S}$ in the former $x$-range, and the linear derivative 
$F_{2,S}^{\,\prime}$ in the latter region. Note that the positive NNLO 
gluon contribution reaches 5\% of the total $|\dot{F}_{2,S}|$ at 
$x = 0.5$, enough to jeopardize purely non-singlet analyses of 
$F_2^{\, p}$ data also in the region $x > 0.3$.

\vspace{1mm}
The reduced $\mu_r$-dependence of both $F_2$ and its derivatives leads
to a better theoretical accuracy of determinations of the parton 
densities from data on $F_{2,S}$ and $dF_{2,S} / d\ln Q^2$ at $Q^2
\simeq 30\mbox{ GeV}^2$: NNLO uncertainties of less than 2\% from 
the truncation of the perturbation series are obtained for the quark
density at $10^{-3}\! < x\! < 0.5$ and for the gluon density
at $3\cdot 10^{-3}\! < x\! < 0.2$.

\section{Summary}

We have briefly discussed the evolution of unpolarized parton densities 
and structure functions in the $\overline{\mbox{MS}}$ scheme. Our 
approximate results for the 3-loop splitting functions $P^{(2)}(x)$ 
pave the way for promoting, even though only at $x\! > \! 10^{-3}$, 
global analyses of DIS and related processes to NNLO accuracy. We will 
also provide approximations for the 3-loop non-singlet coefficient 
functions $c_{\rm NS}^{(3)}$, thus enabling NNNLO determinations of 
$\alpha_s$ from structure functions at least at $x\!\geq \! 0.2$.

\vspace{1mm}
At very large $x$, $x \gsim 0.8$, terms even beyond NNNLO are relevant. 
Here results are available from soft-gluon resummation 
\cite{sglue,av99}.  Progress towards the important HERA small-$x$ 
region of $x \lsim 10^{-3}$ at moderate/low $Q^2$ requires the full 
calculation of the three-loop splitting functions~\cite{MV2}.

\vspace{1mm}
{\sc Fortran} subroutines of our parametrizations of the 2-loop
coefficient functions and our approximations of the 3-loop 
splitting functions can be obtained from neerven@%
lorentz.leidenuniv.nl or avogt@lorentz.leidenuniv.nl.

\end{document}